\documentclass[a4paper,11pt]{article}

\usepackage{jheppub_mod} 
\usepackage[T1]{fontenc} 

\usepackage{hyperref}
\usepackage{graphicx}
\usepackage{amsmath,amssymb,slashed}
\usepackage{booktabs,tabulary}
\usepackage{dsfont}
\usepackage{color}
\usepackage[dvipsnames]{xcolor}
\usepackage{multirow}
\usepackage{tikz}
\usepackage[compat=1.1.0]{tikz-feynman}
\usepackage[compat=1.1.0]{tikz-feynhand}
\usepackage{braket}
\usepackage[capitalise]{cleveref}
\usepackage{bbold}
\usepackage{soul}
\usepackage{physics}

\newcommand{\keV}{\,\text{keV}}

\newcommand{\GeV}{\,\text{GeV}}
\newcommand{\F}{\mathcal{F}}

\newcommand{\etap}{\eta^{(\prime)}}

\newcommand{\thetaSO}{\theta_{81}}
\newcommand{\phiFL}{\phi_{qs}}
\newcommand{\phiKin}{\phi_{\pi\ell}}

\allowdisplaybreaks[1]

\title{\boldmath{$CP$ violation in $\eta^{(\prime)}\to\pi^+\pi^-\mu^+\mu^-$ decays}}

\author[a,b]{Maximilian Zillinger,}
\author[a]{Bastian Kubis,}
\author[c,d]{and Pablo Sánchez-Puertas}

\affiliation[a]{
Helmholtz-Institut f\"ur Strahlen- und Kernphysik (Theorie) and \\
Bethe Center for Theoretical Physics, Universit\"at Bonn, 53115 Bonn, Germany}

\affiliation[b]{Albert Einstein Center for Fundamental Physics, Institute for Theoretical Physics,\\ University of Bern, Sidlerstrasse 5, 3012 Bern, Switzerland}

\affiliation[c]{Institut de Física d’Altes Energies (IFAE) and Barcelona Institute of Science and Technology (BIST),
Campus UAB, 08193 Bellaterra (Barcelona), Spain}
\affiliation[d]{Departamento de F{\'i}sica At{\'o}mica, Molecular y Nuclear, Universidad de Granada, 18071 Granada, Spain}

\emailAdd{zillinger@itp.unibe.ch}
\emailAdd{kubis@hiskp.uni-bonn.de}
\emailAdd{pablosanchez@ugr.es}

\abstract{
It has been pointed out recently that a certain set of dimension-6 scalar $P$- and $CP$-violating light-quark--muon operators may be tested in $\eta$ and $\eta'$ decays to various final states involving $\mu^+\mu^-$ pairs, at a level not yet excluded by constraints from electric dipole moments.  We here work out the hadronic matrix elements required for the predictions for the decays $\eta, \eta' \to\pi^+\pi^-\mu^+\mu^-$.  We relate a new asymmetry in the angular distribution between the dipion and dimuon decay planes to the corresponding Wilson coefficients.  Despite the advantage of not requiring the measurement of muon polarization, the projected sensitivities are shown to be moderate due to a combination of small phase space and chiral suppression.
Such sensitivity studies are most timely in view of upcoming or planned high-statistics $\eta^{(\prime)}$ experiments such as the JLab eta factory or REDTOP.
}

\begin{document} 
\maketitle

\section{Introduction}
\label{sec:intro}

Low-energy high-luminosity experiments are attracting attention due to the absence of physics beyond the Standard Model (BSM) at the LHC that far. In this context, the tests of discrete symmetries offer a suggestive avenue to search for new physics, also in the field of $\eta$ and $\eta'$ physics thanks to the proposed high-luminosity $\etap$ factories such as JEF~\cite{Gan:2017kfr} or REDTOP~\cite{REDTOP:2022slw}. Notwithstanding, $CP$ violation is highly constrained. In particular, $C$-even $P$-odd observables suffer from stringent bounds due to electric dipole moment (EDM) constraints~\cite{Gan:2020aco}. A pertinent theoretical effort is to assess such bounds, which is timely with the forthcoming high-luminosity $\eta$ factories. 

In Refs.~\cite{Sanchez-Puertas:2018tnp,Sanchez-Puertas:2019qwm}, this task was initiated in the context of the Standard Model effective field theory (SMEFT), with focus on leptonic $\etap$ decays. There, it was shown that $CP$ violation rooted in the strongly-interacting or the electromagnetic sector is severely constrained~\cite{Sanchez-Puertas:2018tnp,Gan:2020aco}, whereas a certain class of $D=6$ quark--lepton Fermi operators, whose EDM contribution appears at the two-loop level, could avoid such constraints. In particular, with the current bounds, $CP$ violation could be observed in polarization observables in $\eta\to\mu^+\mu^-$ decays at REDTOP. These kind of operators could leave as well $CP$-violating imprints on semileptonic $\eta$ decays, whose study was relegated as they require a dedicated evaluation of hadronic matrix elements. In Ref.~\cite{Escribano:2022zgm}, the $\etap\to\pi^0\mu^+\mu^-$ (as well as $\eta'\to\eta\mu^+\mu^-$) decays 
were assessed, finding that $CP$-violating signatures in such decays are beyond REDTOP statistics, partly due to isospin-breaking suppression. 

In this work, we perform a dedicated study of $\eta^{(\prime)}\to\pi^+\pi^-\mu^+\mu^-$ decays. Compared to Ref.~\cite{Escribano:2022zgm}, the $CP$-violating new-physics contribution is free of an isospin-breaking suppression and its $CP$-violating signature does not require muon polarimetry, which makes this an \textit{a priori} interesting case. We note that such a process has been considered long ago in the context of $CP$ violation~\cite{Gao:2002gq} (see also Ref.~\cite{Geng:2002ua}), albeit focusing on $CP$-violating electromagnetic form factors that can be excluded based on EDM constraints~\cite{Gan:2020aco}. 
As a result, we show that in the current context a new plane asymmetry should be considered in $CP$-violating searches. Further, we find that, within the SMEFT framework, $CP$-violating signatures in $\eta^{(\prime)}\to\pi^+\pi^-\mu^+\mu^-$ decays are tightly constrained by the nEDM at foreseeable future $\etap$ factories, with $\eta\to\mu^+\mu^-$ still the most promising case.

This article is organized as follows. In \cref{sec:main}, we introduce the Standard-Model (SM) process as well as the SMEFT $CP$-violating contribution, deriving the relevant plane asymmetry that has been overlooked so far. Such an asymmetry requires the relevant matrix elements for the SM and $CP$-violating contributions as input, which are described in \cref{sec:SM,sec:BSM}, respectively, with the latter representing the major work in this study. Our final results and conclusions are given in \cref{sec:results,sec:summary}.

\section{Decay amplitude and \texorpdfstring{$CP$}{CP}-violating observables}
\label{sec:main}

In the Standard Model (SM), at leading order, the $\eta^{(\prime)}(P)\to\pi^+(p_1)\pi^-(p_2)\mu^+(p_3)\mu^-(p_4)$ decay amplitude is both $C$- and $P$-even and reads
\begin{equation}\label{Eq:SMmatrixelement}
 \mathcal{M}_{\textrm{SM}} = \frac{-e^2}{s_{\ell}}[\bar{u}(p_4)\gamma^{\mu} v(p_3)]
 \big\langle \pi^+(p_1)\pi^-(p_2) 
 \big|  j_{\mu}(0) \big| 
 \eta^{(\prime)}(P) \big\rangle  , 
\end{equation}
with $j_{\mu} = (2/3)\bar{u}\gamma_{\mu}u - (1/3)\bar{d}\gamma_{\mu}d - (1/3)\bar{s}\gamma_{\mu}s$. The non-perturbative hadronic matrix element above can be expressed as~\cite{Gan:2020aco}
\begin{equation}
    \big\langle\pi^+(p_1)\pi^-(p_2) 
    \big| j_{\mu}(0) \big| 
    \eta^{(\prime)}(P)\big\rangle   = \epsilon_{\mu\nu\alpha\beta} p_1^{\nu}p_2^{\alpha}q^{\beta}\F_{\eta^{(\prime)}}(s,t,u;s_{\ell}),
\end{equation}
where $\epsilon^{0123}=1$, $s=(p_1 +p_2)^2$, $t=(P -p_1)^2$, $u=(P -p_2)^2$, $q =p_3 +p_4$, and $s_{\ell}=(p_3+p_4)^2$, and is discussed in \cref{sec:SM}.   

In the presence of new physics, additional $CP$-violating contributions might appear. In the following, we adopt SMEFT to parameterize them, implicitly assuming heavy new physics. In this context, there are different possible sources of $CP$ violation. One possibility, studied long ago~\cite{Gao:2002gq},\footnote{Introducing  $p= p_{\pi^+} +p_{\pi^-}$ and $\bar{p}= p_{\pi^+} -p_{\pi^-}$, the most general $CP$-violating hadronic matrix element in \cref{Eq:SMmatrixelement} can be written as $ (A_1/2)[p^{\mu}(\bar{p}q) -(pq)\bar{p}^{\mu} ] +A_2[q^2 \bar{p}^{\mu} - (\bar{p}q)q^{\mu}]$, where $q$ labels the photon momentum, which can be derived from $\mathcal{L}_{\textrm{eff}} = -i A_1 \eta F^{\mu\nu} \partial_{\mu}\pi^+\partial_{\nu}\pi^- + iA_2 \eta \partial_{\mu}F^{\mu\nu}(\pi^+\partial_{\nu}\pi^- - \pi^-\partial_{\nu}\pi^+)$. The one in Ref.~\cite{Gao:2002gq} corresponds to the $A_1$ term, which is bounded by the neutron's EDM~\cite{Gan:2020aco}. 
In turn, using $\partial_{\mu}F^{\mu\nu} = -ej^{\nu}$ the second term can be conceived microscopically as a $(\bar{\ell}\gamma^{\mu}\ell)(\mathcal{O}_{\mu})$ hadron--lepton interaction, where $\mathcal{O}_{\mu}$ is an operator built from quark (and possibly gluon) fields. Note however that, based on discrete symmetries, such a structure requires $D>3$ to hadronize into  $\eta(\pi^+\partial_{\nu}\pi^- - \pi^-\partial_{\nu}\pi^+)$. Its possible origin in a SMEFT language is thus of $D>6$, and \textit{a priori} less relevant than the $D=6$ case here discussed.} consists on $CP$-violating electromagnetic form factors, whose origin in SMEFT might be in the QCD sector or quark EDMs. Regardless of its origin, the neutron's EDM (nEDM) sets stringent bounds on such a scenario~\cite{Gan:2020aco}. Consequently, following Ref.~\cite{Sanchez-Puertas:2018tnp}, we shall focus on quark--lepton Fermi operators arising at $D=6$ in SMEFT. Compared to Refs.~\cite{Sanchez-Puertas:2018tnp,Escribano:2022zgm}, besides scalar operators, tensor operators can contribute as well. However, these produce a non-vanishing nEDM at the one-loop level, and are tightly constrained, see Ref.~\cite{Cirigliano:2017tqn}. For this reason, we will focus on the scalar operators studied in Refs.~\cite{Sanchez-Puertas:2018tnp,Escribano:2022zgm}, whose contribution to the nEDM appears at the two-loop level, lessening the nEDM bounds. In particular, we will restrict ourselves to the flavor-neutral $CP$-violating part of the $\mathcal{O}_{\ell equ}^{(1)}$ and $\mathcal{O}_{\ell edq}$ operators~\cite{Grzadkowski:2010es} and will consider the muonic case only, since electrons suffer from more stringent bounds from atomic physics~\cite{Yanase:2018qqq} (see also comments on the additional chiral suppression below).
We pay special attention to the contribution of the strange-quark operator (absent in $\mathcal{O}_{\ell equ}^{(1)}$), which is again less constrained from the nEDM than those of the first quark generation~\cite{Sanchez-Puertas:2018tnp}.  The operators in question read 
\begin{align}
\mathcal{O}_{\ell equ}^{(1)} & \to 
    -\frac{G_F}{\sqrt{2}} \operatorname{Im} c_{\ell equ}^{(1)2211}
    \big[ (\bar{\mu}i\gamma^5\mu)(\bar{u} u)
    +(\bar{\mu}\mu)(\bar{u} i\gamma^5 u)\big], \notag\\
\mathcal{O}_{\ell edq} & \to 
    \frac{G_F}{\sqrt{2}} \Big\{ \operatorname{Im} c_{\ell edq}^{2211}
    [ (\bar{\mu}i\gamma^5\mu)(\bar{d} d)
    -(\bar{\mu}\mu)(\bar{d} i\gamma^5 d)] \notag\\
    &\qquad\quad  +
    \operatorname{Im} c_{\ell edq}^{2222}
    [ (\bar{\mu}i\gamma^5\mu)(\bar{s} s)
    -(\bar{\mu}\mu)(\bar{s} i\gamma^5 s)]
    \Big\}
    , 
\end{align}
which produce the following $CP$-violating contribution to the current process: 
\begin{align}
    \mathcal{M}_{\textrm{BSM}} &=  
    - \frac{G_F}{2\sqrt{2}}  
    \operatorname{Im}\big(c_{\ell e qu}^{(1)2211}+c_{\ell edq}^{2211}\big)
    [\bar{u}(p_4)v(p_3)] \big\langle \pi^+(p_1)\pi^-(p_2) \big| P^q \big| \eta^{(\prime)}(P) \big\rangle 
    \notag\\
    &- \frac{G_F}{\sqrt{2}} \operatorname{Im}c_{\ell edq}^{2222}
    [\bar{u}(p_4)v(p_3)] \big\langle \pi^+(p_1)\pi^-(p_2) \big| P^s \big| \eta^{(\prime)}(P) \big\rangle ,
\end{align}
with the pseudoscalar operators $P^{q,s}=\{ \bar{u}i\gamma^5 u +\bar{d}i\gamma^5 d, \bar{s}i\gamma^5 s \}$ for the light- and strange-quark contribution, respectively.
The combination of the Wilson coefficients in the first line follows from isospin symmetry.
Note that all Wilson coefficients depend on the renormalization scale, as the (pseudo)scalar quark operators are not renormalization group invariant.
The description of the hadronic matrix elements above is postponed to \cref{sec:BSM},  as it is inessential for the following discussion concerning the $CP$-violating asymmetry. Such an asymmetry is produced via the interference of the SM and BSM contributions due to their opposite $CP$ nature,
\begin{equation}
    d\Gamma = \frac{d\Pi_4}{2M_{\eta^{(\prime)}}} |\mathcal{M}|^2 \simeq 
    \frac{d\Pi_4}{2M_{\eta^{(\prime)}}} \left[ |\mathcal{M}_{\textrm{SM}}|^2 + 2\operatorname{Re}\mathcal{M}_{\textrm{SM}}^*\mathcal{M}_{\textrm{BSM}} + |\mathcal{M}_{\textrm{BSM}}|^2 \right],
\end{equation}
with $d\Pi_4$ standing for the four-body phase space, see \cref{sec:AppPhaseSpace}. Among the different terms, only the interference one is of $CP$-odd nature. Making use of the definitions above, one finds that 
\begin{align}
    2\operatorname{Re}\mathcal{M}_{\textrm{SM}}^*\mathcal{M}_{\textrm{BSM}} &= 
    \frac{8\sqrt{2}e^2m_{\mu} G_F }{s_{\ell}} \epsilon_{\mu\nu\rho\sigma}p_1^{\mu}p_2^{\nu}p_3^{\rho}p_4^{\sigma}  \operatorname{Re}\bigg[\F^*_{\eta^{(\prime)}}(s,t,u;s_{\ell}) \notag \\ &
    \times \bigg(
    \big\langle \pi^+\pi^- \big|
    \frac{1}{2}\operatorname{Im} \big(c_{\ell equ}^{(1)2211} +c_{\ell edq}^{2211}\big)P^{q} +\operatorname{Im} c_{\ell edq}^{2222}P^{s}
    \big| \eta^{(\prime)} \big\rangle
    \bigg)
    \bigg].
\end{align}
The interference term above is of $C$-even, $P$-odd nature and, while it does not  contribute to the total decay width, it produces an angular asymmetry among the dilepton and dipion planes. To show this, it is convenient to parameterize the phase space as described in \cref{sec:AppPhaseSpace}, in which the fully-contracted Levi-Civita tensor is proportional to $\sin\phiKin$ (where $\phi_{\pi\ell}$ is the angle between the dilepton and dipion planes), allowing us to introduce the angular asymmetry 
\begin{align} \label{eq:Asinphi}
    A_{\phi}^{\eta^{(\prime)}} &= \frac{N(\phiKin\in[0,\pi]) -N(\phiKin\in[\pi,2\pi])}{N} \notag\\
    &= 
    -\frac{1}{\Gamma}\frac{\sqrt{2} G_F\, \alpha \, m_{\mu}}{2^{12}\pi^4 M_{\eta^{(\prime)}}^3} \int ds\, ds_{\ell}\, d\cos\theta_{\pi}\, \lambda \sqrt{s/s_{\ell}}
    \beta_{\ell}^2\beta_{\pi}^2\sin\theta_{\pi} \notag\\
&\times\operatorname{Re}\Big[\F^{*}_{\eta^{(\prime)}}(s,t,u;s_{\ell})\big\langle \pi^+\pi^- \big|
    \frac{1}{2}\operatorname{Im} \big(c_{\ell equ}^{(1)2211} +c_{\ell edq}^{2211}\big)P^{q} +\operatorname{Im} c_{\ell edq}^{2222}P^{s}
    \big| \eta^{(\prime)} \big\rangle \Big] ,
\end{align}
where $N$ stands for the number of events, $\Gamma$ for the partial width, $2t(2u) = 2M_{\pi}^{2}+M_{\eta^{(\prime)}}^2 -s +s_{\ell} \mp\lambda^{1/2}\beta_{\pi}\cos\theta_{\pi}$, with
$\beta_{\pi}^2 = 1 -4M_\pi^2/s$, 
$\beta_{\ell}^2 = 1 -4m_\ell^2/s_{\ell}$, 
$\theta_{\pi}$ is the angle in the $\pi^+\pi^-$
frame, and $\lambda\equiv\lambda(M_{\eta^{(\prime)}}^2,s,s_{\ell})$, with $\lambda(a,b,c)=a^2+b^2+c^2-2ab-2ac-2bc$ the K{\"a}ll{\'e}n function. 
Note in addition the chiral suppression $m_{\mu}/M_{\eta}$ due to the different chiral structure of the SM and BSM contributions, which would imply a huge suppression for the electronic case, making it even more irrelevant for our purpose, even beyond the stronger bounds from atomic EDMs.
With the hadronic input at hand, the equation above allows us to estimate the sensitivity of $A_{\phi}$ to the Wilson coefficient, which is a valuable information for experimentalists. Importantly, it is common in experimental searches to look for a $\sin(2\phi)$ asymmetry instead~\cite{KLOE:2008arm}. Such searches would miss the signature above, so an important message for the experimentalists is to incorporate the $A_{\phi}$ asymmetry in their analyses.  We point out that in the similar decay $K_L\to\pi^+\pi^- e^+e^-$, the (dominant) $\sin(2\phi)$ asymmetry is associated with indirect $CP$ violation, while a $\sin\phi$ asymmetry would indicate direct $CP$ violation~\cite{Heiliger:1993qt,Ecker:2000nj}. 

The following sections are dedicated to the description of the hadronic matrix elements that feed the equation above, with focus on the BSM part, whose corresponding matrix element has not been assessed earlier to the best of our knowledge.


\section{Standard-Model decay amplitudes}
\label{sec:SM}

In this section, we briefly describe the hadronic matrix elements $\F_{\eta^{(\prime)}}(s,t,u;s_{\ell})$. 
At vanishing energies and in the chiral limit, they are given by low-energy theorems due to the Wess--Zumino--Witten anomaly~\cite{Wess:1971yu,Witten:1983tw},
\begin{equation}
    \F_{\eta}(0,0,0;0) = \frac{1}{4\sqrt{3}\pi^2F_\pi^3} \approx 18.6\GeV^{-3} , \qquad
    \F_{\eta'}(0,0,0;0) = \frac{1}{2\sqrt{6}\pi^2F_\pi^3} \approx 26.3\GeV^{-3},
\end{equation}
where for simplicity, we have also adopted the large-$N_c$ limit in order to avoid further corrections to the singlet part.  
Beyond that limit, these decays have been studied in chiral perturbation theory ($\chi$PT) for real photons~\cite{Bijnens:1989ff}, as well as in unitarized versions thereof both for real and virtual photons~\cite{Borasoy:2004qj,Borasoy:2007dw}.  More recently, a model-independent dispersion-theoretical description in terms of the universal pion--pion final-state interactions has been suggested~\cite{Stollenwerk:2011zz}, which reduces the description of the $\pi\pi$ spectra to a few phenomenological polynomial parameters, which can be either matched to $\chi$PT or fitted to experimental data~\cite{WASA-at-COSY:2011fzp,KLOE:2012rfx,BESIII:2017kyd}.

Following  Refs.~\cite{Kubis:2015sga,Hanhart:2016pcd,Holz:2015tcg,Holz:2022hwz}, the decay amplitudes can be expanded in partial waves according to 
\begin{equation}
    \F_{\eta^{(\prime)}}(s,t,u;s_{\ell}) = \sum_{\textrm{odd}~l} P_l^{\prime}(\cos\theta_\pi) f_{l}^{\eta^{(\prime)}} \!(s,s_{\ell}), \qquad 
    \cos\theta_\pi =  \frac{t-u}{ \lambda^{1/2}(M_{\eta^{(\prime)}}^2, s,s_{\ell}) \beta_\pi(s)}.
\end{equation}
Since the amplitude is completely dominated by the first partial wave $l=1$ in the decay region~\cite{Kubis:2015sga}, we adopt the $P$-wave approximation $\F_{\eta^{(\prime)}}(s,t,u,s_{\ell}) = f_1^{\eta^{(\prime)}}\!(s,s_{\ell})$. Further, we shall make use of a model based on a factorization ansatz for the dependence on $s$ and $s_l$, which has been shown to work surprisingly well even for dilepton energies in the range $1\GeV\leq\sqrt{s_l}\leq2\GeV$~\cite{Holz:2015tcg}, and is therefore expected to be safely sufficient to very high accuracy in the decay region(s) of interest here. This simplifies the low-energy description, which can be expressed in terms of the well-known Omn{\`e}s function~\cite{Omnes:1958hv} as
\begin{equation}
    f_{1}^{\eta^{(\prime)}}\!(s,s_{\ell}) = f_1^{\eta^{(\prime)}}\!(s) \bar{F}(s_{\ell}) = P_{\eta^{(\prime)}}(s)\Omega_1^1(s)\bar{F}(s_{\ell}), \quad 
    \Omega_1^1(s) = \operatorname{exp}\left( \frac{s}{\pi} \int dz \frac{\delta_1^1(z)}{z(z-s)} \right),
\end{equation}
with $P_{\eta^{(\prime)}}(s)$ a polynomial, $\delta_1^1$ the $P$-wave $\pi\pi$ phase shift~\cite{GarciaMartin:2011cn}, and $\bar{F}(0) = \Omega_1^1(0) = 1$  such that $f_{1}^{\eta^{(\prime)}}\!(0,0) = P_{\eta^{(\prime)}}(0)$. 
For $P_{\eta^{(\prime)}}(s)$, we follow the work in Refs.~\cite{Stollenwerk:2011zz,Hanhart:2016pcd,Holz:2022hwz}, where the following parameterizations were used:
\begin{align}
  P_{\eta}(s) = A_{\eta}(1 +\alpha_{\eta}s), \qquad 
  P_{\eta'}(s) = A_{\eta'} \left(1 +\alpha_{\eta'}s +\beta_{\eta'}s^2 +\frac{\kappa_2}{M_{\omega}^2 -s -iM_{\omega}\Gamma_{\omega}} \right), 
\end{align}
with input taken from their analyses of $\eta^{(\prime)}\to\pi^+\pi^-\gamma$ decays. 
For the decay of the $\eta$ meson it was shown that a linear polynomial is sufficient to describe data by the KLOE collaboration in the physical decay region~\cite{KLOE:2012rfx}, yielding 
$A_{\eta}=(17.9\pm 0.4\mp 0.1)\, \text{GeV}^{-3}$ and $\alpha_{\eta}=(1.52\pm0.06)\,\textrm{GeV}^{-2}$~\cite{Kubis:2015sga}. In the case of the $\eta^{\prime}$ decay, a quadratic polynomial seems to fit the decay data very accurately, with the exception of isospin-breaking $\rho$--$\omega$ mixing effects. The values of the coefficients are obtained by a fitting procedure to the BESIII data~\cite{BESIII:2017kyd}, resulting in 
$A_{\eta^{\prime}}=16.7(4)\, \text{GeV}^{-3}$, $\alpha_{\eta^{\prime}}=1.00(4)\,\textrm{GeV}^{-2}$, $\beta_{\eta^{\prime}}=-0.55(4)\,\text{GeV}^{-4}$, and $\kappa_2=6.72(24)\times 10^{-3}\,\text{GeV}^{-1}$~\cite{Hanhart:2016pcd}. Concerning $\bar{F}(s_l)$ note that, besides our convention $\bar{F}(0)=1$, it should only contain $I=1$ vector resonances. As such, we adopt the following model for the form factor 
\begin{align}
    \bar{F}(s_{\ell}) &= \frac{M_{\rho}^2 M_{\rho'}^2}{ [M_{\rho}^2 -s_{\ell} -i\sqrt{s_{\ell}}\Gamma_{\rho}(s_{\ell})] [M_{\rho'}^2 -s_{\ell} -i\sqrt{s_{\ell}}\Gamma_{\rho'}(s_{\ell})] }, \notag\\ 
    \Gamma_V(s_{\ell}) &= \Gamma_V 
    \frac{\sqrt{s_{\ell}}}{M_V}\frac{\beta_{\pi}^3(s_{\ell})}{\beta_{\pi}^3(M_V^2)}    
    \theta(s_{\ell} -4M_{\pi}^2),
\end{align}
with $\beta_{\pi}(s_{\ell}) = (1-4M_{\pi}^2/s_{\ell})^{1/2}$, that in addition ensures the appropriate asymptotic behavior in the kinematic region of interest. To see this, note that the $\pi^+\pi^-$ landscape is fully dominated by the $\rho$ resonance at the energies of interest, so that our form factor can be effectively thought of as $\bra{\pi^+\pi^-} j_{\mu}(0) \ket{\eta} \to \bra{\rho} j_{\mu}(0) \ket{\eta}$, with the latter behaving asymptotically as $1/q^4$~\cite{Chernyak:1983ej,Zhang:2015mxa}. With the current description we obtain $\Gamma(\eta\to\pi^+\pi^-\mu^+\mu^-) = 8.33(40)\times10^{-15}\,\textrm{GeV}$ and $\Gamma(\eta'\to\pi^+\pi^-\mu^+\mu^-) = 4.23(24)\times10^{-9}\,\textrm{GeV}$, which will be used in \cref{sec:results}. 
The resulting branching fractions are compared to experiment in \cref{tab:SM-BRs}, where for completeness we also include the numbers for the electron--positron final states.  We find excellent agreement throughout, which further supports our description.

\begin{table}
	\centering
	\def\arraystretch{1.3}
	\begin{tabular}{cccc}
		\toprule
		Channel & This work & Experiment & Ref.   \\
		\midrule 
		$\eta\to\pi^+\pi^-e^+e^-$ & $2.65(17)\times10^{-4}$ & $2.68 (9)_\text{stat.}(7)_\text{syst.} \times 10^{-4}$ & \cite{KLOE:2008arm} \\
		$\eta\to\pi^+\pi^-\mu^+\mu^-$ & $6.36(39)\times 10^{-9}$ & --- & \\
		$\eta'\to\pi^+\pi^-e^+e^-$ & $2.21(15)\times10^{-3}$& $2.11(12)_\text{stat.}(15)_\text{syst.} \times10^{-3}$& \cite{BESIII:2013tjj} \\
		$\eta'\to\pi^+\pi^-\mu^+\mu^-$ & $2.25(14)\times 10^{-5}$ & $1.97(33)_\text{stat.}(19)_\text{syst.}\times 10^{-5}$ & \cite{BESIII:2020elh} 
		\\ \bottomrule
	\end{tabular}
\caption{Branching ratios for the various $\etap\to\pi^+\pi^-\ell^+\ell^-$ decay channels in the Standard Model.}
\label{tab:SM-BRs}
\end{table}


\section{Beyond-the-Standard-Model decay amplitudes}
\label{sec:BSM}

The main task in our work is the evaluation of the pseudoscalar matrix elements that enter the BSM contribution. This is split into the $\eta$ and $\eta'$ parts, as they require using different frameworks. In particular, for the $\eta$ case we shall make use of $SU(3)$ $\chi$PT. We advocate that, while for the light-quark case an LO calculation should provide us with a reasonable estimate, the strange-quark case demands an NLO evaluation due to the strong LO chiral suppression, which is due to soft-pion theorems~\cite{deAlfaro:1973zz}. 
For the $\eta'$, we shall employ large-$N_c$ $\chi$PT at NLO (which is important due to the large $\eta'$ mass) together with a unitarization method for the final $\pi^+\pi^-$ state.

\subsection{BSM \texorpdfstring{$\eta$}{eta} matrix elements}
\label{sec:etaBSM}

The current matrix elements $\bra{\pi^+ \pi^-}  P^a \ket{\eta}$
have not been computed before to the best of our knowledge and might be of interest for model of new physics featuring pseudoscalar couplings. 
At LO, there are two Feynman diagrams from the leading chiral Lagrangian: those where the pseudoscalar current produces an $\eta$ meson that later couples to an  $\eta\pi^+\pi^-$ state, and  those where the three pseudoscalar mesons are directly sourced from the pseudoscalar current, see \cref{Fig:LOeta}.
\begin{figure}
	\centering
	\scalebox{1.0}{
		\begin{tikzpicture}[baseline=(b)]
			\begin{feynman}[large]
				\vertex (a) {$\eta$};
				\vertex [right=2.0cm of a, dot] (b) {};
				\vertex [above=0.5cm of b] (g) {};
				\vertex [above right=2.0cm of b] (c) {$\pi^{+}$};
				\vertex [right=2.0cm of b] (d) {$\pi^{-}$};
				\vertex [below right=2.0cm of b, crossed dot] (e) {};
				\vertex [below right=2.7cm of b] (f) {$P^a$};
			    \diagram*{
					(a)  --  (b),
					(b) -- (c),
					(b) -- (d),
					(b) -- [edge label'=\( \eta \)] (e),
				};
			\end{feynman}
		\end{tikzpicture}
	}
    \hspace{1.5cm}
	\scalebox{1.0}{
		\begin{tikzpicture}[baseline=(b)]
			\begin{feynman}[large]
				\vertex (a) {$\eta$};
				\vertex [right=2.0cm of a, crossed dot] (b) {};
				\vertex [right=1.8cm of a] (h);
				\vertex [below=0.2cm of h] (c) {$P^a$};
				\vertex [above right=2.0cm of b] (d) {$\pi^{+}$};
				\vertex [below right=2.0cm of b] (e) {$\pi^{-}$};
				\diagram*{
					(a)  --  (b),
					(b) -- (d),
					(b) -- (e),
				};
			\end{feynman}
		\end{tikzpicture}
	}
	\caption{Leading-order (LO) Feynman diagrams for $\matrixel{\pi^{+}\pi^{-}}{P^a}{\eta}$ in $\chi$PT. 
	}
	\label{Fig:LOeta}
\end{figure}
As such, it is natural to separate the amplitude even beyond LO according to
\begin{equation}
    \bra{\pi^+\pi^-} P^a \ket{\eta}
    = \bra{0} P^a \ket{\eta}\frac{1}{M_{\eta}^2 -s_\ell} \mathcal{M}_{\eta\eta\to\pi\pi} + \mathcal{M}^{\eta;a}_{\textrm{non-pole}}, \label{eq:EtaAmp}
\end{equation}
where the residues of the $\eta$ poles are given by the pseudoscalar matrix element of the $\eta$ with appropriate flavor, multiplied by the on-shell $\eta\eta\to\pi\pi$ scattering amplitude $\mathcal{M}_{\eta\eta\to\pi\pi}$.  
In this respect, it is important to note that the separation in Eq.~\eqref{eq:EtaAmp} is not necessarily in one-to-one correspondence to specific Feynman diagrams, which are in general representation dependent: $\mathcal{M}^{\eta}_{\textrm{non-pole}}$ also receives contributions from diagrams where the pseudoscalar current couples to an $\eta$ meson and off-shell effects need to be accounted for. 
At NLO, this prevents us from taking the standard results for the $\mathcal{M}_{\eta\eta\to\pi\pi}$ scattering amplitude~\cite{Bernard:1991xb,GomezNicola:2001as,Albaladejo:2015aca} straight away.

At LO, the two matrix elements in question are given by
\begin{align}
\bra{\pi^+(p_1) \pi^-(p_2)}  P^s \ket{\eta(P)} &=
-\frac{2B_0}{3\sqrt{3}F_{\pi}} \frac{M_{\pi}^2}{M_{\eta}^2-s_\ell} ,
\notag\\
    \bra{\pi^+(p_1) \pi^-(p_2)}  P^q  \ket{\eta(P)} &= 
    -\frac{2B_0}{\sqrt{3}F_{\pi}}
    \left(  
       1 -\frac{M_{\pi}^2}{3(M_{\eta}^2-s_\ell)}
    \right) . \label{eq:etaQMatEl}
\end{align}
The low-energy constant $B_0$ that is related to the quark condensate appears via the pseudoscalar matrix element $\bra{0} P^a \ket{\eta}$ and the non-pole term $\mathcal{M}^{\eta;a}_{\textrm{non-pole}}$. Throughout this work we use the value $B_{0}=2.39(18)\,\textrm{GeV}$~\cite{MILC:2009ltw,Aoki:2019cca}, valid at a scale of $2\GeV$.  Scale invariance is restored in the products of $B_0$ with the corresponding Wilson coefficients.  We note that, in both cases, the $\eta$-pole contribution is chirally suppressed by $M_\pi^2$, which is a result of the corresponding suppression of the LO scattering amplitude $\mathcal{M}_{\eta\eta\to\pi\pi}^{\textrm{LO}} =M_{\pi}^2/(3F_{\pi}^2)$.  The residue is fixed completely by 
$\bra{0} P^q \ket{\eta} = - \bra{0} P^s \ket{\eta} = 2F_{\pi} B_0/\sqrt{3}$.  Furthermore, for the strange current, there is no non-pole contribution at all, such that the chiral suppression actually holds for the complete matrix element at LO. 
As a result, while the LO result can provide a reasonable estimate for the light-quark matrix element, the strange one receives 
comparatively large NLO corrections, which we evaluate in the following. First,
to render the separation into non-pole and pole parts unambiguous at NLO,\footnote{Note that $s+t+u -2M_{\pi}^2 -2M_{\eta}^2 =  (s_{\ell} -M_{\eta}^2)$ can be used to modify the pole and non-pole parts.} we shall employ the decomposition~\cite{Albaladejo:2015aca}
\begin{align}
    \mathcal{M}_{\eta\eta\to\pi\pi}(s,t,u) &= \frac{M_{\pi}^2}{3F_{\pi}^2} 
    +W_{\eta\eta}(s) +U_{\eta\eta}(t) +U_{\eta\eta}(u), \label{eq:PoleDeco} \\
    \mathcal{M}^{\eta;s}_{\textrm{non-pole}}(s,t,u;s_\ell) &= W^{\eta;s}_{\textrm{non-pole}}(s,s_\ell) +U^{\eta;s}_{\textrm{non-pole}}(t) +U^{\eta;s}_{\textrm{non-pole}}(u). \label{eq:NonPoleDeco}
\end{align}
We follow Ref.~\cite{Isken:2017dkw} and impose $U_{\eta}(0) = U^{\prime}_{\eta}(0)= 0$, which uniquely fixes the separation at NLO.
The full NLO results for the quantities in Eqs.~\eqref{eq:EtaAmp}, \eqref{eq:PoleDeco}, and \eqref{eq:NonPoleDeco} are collected in \cref{sec:AppBSMeta}. Note there the presence of contributions of the kind $L_i M_{\eta}^4/F_{\pi}^4$ to $\mathcal{M}_{\eta\eta\to\pi\pi}$, which have a size that is comparable to the LO result for $\mathcal{M}_{\eta\eta\to\pi\pi}^{\textrm{LO}}$. The decomposition adopted therein is such that each individual part in \cref{eq:EtaAmp} is UV-finite and scale-independent, and the scattering amplitude reduces to that in Refs.~\cite{Bernard:1991xb,GomezNicola:2001as,Albaladejo:2015aca} on-shell, providing a cross-check of our calculation. 

A final cross-check comes from  the following soft-pion theorem: 
\begin{equation}\label{eq:softpion}
    \lim_{\substack{p_{\pi^c}\to 0\\ M_{\pi}^2\to0}} -F_{\pi} \big\langle \pi^b\pi^c \big| \bar{q} i\gamma^5 \lambda^a q  \big| \eta \big\rangle = 
    \lim_{M_{\pi}^2\to0} \frac{1}{2}
    \big\langle \pi^b \big| \bar{q}\{ \lambda^a, \lambda^c \} q \big| \eta \big\rangle,
\end{equation}
where $\lambda^c$ fulfills $\bra{0} A_{\mu}^c \ket{\pi^c(p)}= ip_{\mu}F_{\pi}$. 
In particular, for the strange-quark operator $\{\lambda^s,\lambda^c\} = 0$ and the amplitude must vanish in such a limit, which corresponds to $M_{\pi}=0$, $s=0$, $t=M_{\eta}^2$, and $u=s_{\ell}$ (as well as $t\leftrightarrow u$). These Adler zeros~\cite{Adler:1965a,Adler:1965b} provide a non-trivial consistency check between pole and non-pole part of the matrix element at NLO (as well as explaining the suppression observed at LO). Furthermore, for $a=q$ the soft-pion theorem relates our calculation to the corresponding $\eta\to\pi$ isovector scalar form factor, which we checked to hold at LO in $\chi$PT.

\subsection{BSM \texorpdfstring{$\eta'$}{eta'} matrix elements}
\label{sec:etapBSM}

To compute the $\eta'$ matrix elements we advocate the use of large-$N_c$ $\chi$PT~\cite{Kaiser:2000gs,Herrera-Siklody:1996tqr,Herrera-Siklody:1997pgy,Guo:2015xva,Bickert:2016fgy}, which combines the large-$N_c$ limit of QCD with $\chi$PT to incorporate the $\eta'$ as the ninth Goldstone boson. Such computation is performed in \cref{sec:etalNc} and parallels that of the $\eta$ case, but with the absence of loops and a few low-energy constants (LECs) that are suppressed in the large-$N_c$ limit, as well as the inclusion of the $\eta-\eta'$ mixing. However, in contrast to the $\eta$ case, an NLO evaluation is mandatory for both matrix elements due to potential $\mathcal{O}(p_{\eta'}^2/\Lambda^2)$ corrections that prove to be significant. Further, the large available phase space demands to include final-state interactions from the $\pi^+\pi^-$ system, which is unitarized in \cref{sec:unit}.

\subsubsection{Large-\texorpdfstring{$N_c$}{Nc} \texorpdfstring{$\chi$PT}{ChPT} amplitudes at NLO\label{sec:etalNc}}

Once more, the matrix element can be expressed as 
\begin{equation}
    \bra{\pi^+\pi^-} P^a \ket{\eta^{\prime}} 
    = 
    \bra{0} P^a \ket{\eta}\frac{\mathcal{M}_{\eta\eta'\to\pi\pi}}{M_{\eta}^2 -s_\ell}  
    +\bra{0} P^a \ket{\eta'}\frac{\mathcal{M}_{\eta'\eta'\to\pi\pi}}{M_{\eta'}^2 -s_\ell}  
    +\mathcal{M}^{\eta';a}_{\textrm{non-pole}}. \label{eq:EtaPAmp}
\end{equation}
Following the notation in \cref{sec:etaBSM}, we decompose the four-meson amplitudes  as
\begin{equation}\label{eq:etapiscatt}
    \mathcal{M}_{AB\to\pi\pi}(s,t,u)  = 
    c_{AB} \frac{M_{\pi}^2}{3F_{\pi}^2} +W_{AB}(s) +U_{AB}(t) +U_{AB}(u), 
\end{equation}
where $c_{AB}$ can be expressed in terms of the octet--singlet (quark-flavor) mixing angles $\thetaSO$ ($\phiFL$) at NLO accuracy~\cite{Escribano:2010wt} as 
\begin{equation} 
c_{\eta\eta} = 3\cos^2\phiFL, \qquad
c_{\eta'\eta} = 3\cos\phiFL\sin\phiFL, \qquad 
c_{\eta'\eta'} = 3\sin^2\phiFL , \label{eq:cee}
\end{equation} 
with
$\sqrt{3}\cos\phiFL= \cos\thetaSO  -\sqrt{2}\sin\thetaSO$ and $\sqrt{3}\sin\phiFL= \sin\thetaSO  +\sqrt{2}\cos\thetaSO$. The value for the octet--singlet mixing angle $\thetaSO$ is taken from the lattice~\cite{largenclecs} 
\begin{equation}
\thetaSO=\dfrac{\theta_0+\theta_8}{2}=\dfrac{-8.1(1.8)^{\circ}-25.8(2.3)^{\circ}}{2}=-17.0(1.5)^{\circ},
\end{equation}
which corresponds to a mixing angle in the quark-flavor basis of $\phiFL=37.8(1.5)^{\circ}$.
Note that the coefficients in Eq.~\eqref{eq:cee} reflect the fact that, except for OZI-rule-violating contributions, only the light-quark content of the $\eta-\eta'$ system contributes to the scattering amplitude. 
The full results are given by
\begin{align}
    W_{AB}(s) =  c_{AB} \bigg[&\frac{4(3L_2 +L_3)}{3F_{\pi}^4}\big( s^2 -2M_{\pi}^4 -M_A^4 -M_B^4\big) \notag \\ &-\frac{4M_{\pi}^2 L_5}{3F_{\pi}^4}\big(2M_{\pi}^2 +M_A^2 +M_B^2\big) +\frac{16M_{\pi}^4 L_8}{F_{\pi}^4} \bigg] +\tilde\Lambda_{AB},
\end{align}
where 
\begin{equation}
   \{ \tilde{\Lambda}_{\eta\eta},  \tilde{\Lambda}_{\eta'\eta'}, \tilde{\Lambda}_{\eta\eta'}\} = 
   \frac{\sqrt{2}M_{\pi}^2\tilde{\Lambda}}{\sqrt{3}F_{\pi}^2}\left\{ 
   \cos\phiFL\sin\thetaSO, 
   -\sin\phiFL\cos\thetaSO,
   \frac{\sin2\thetaSO -\sin2\phiFL}{2}
   \right\},
\end{equation}
with $\tilde{\Lambda} = \Lambda_1 -2\Lambda_2$~\cite{Kaiser:2000gs,Bickert:2016fgy} a scale-independent combination of OZI-rule-violating parameters. We take $\tilde{\Lambda} =-0.51(21)$ from the lattice~\cite{largenclecs}. Similarly, 
\begin{equation}
    U_{AB}(t) = c_{AB}\frac{4(3L_2 +L_3)}{3F_{\pi}^4}t^2.
\end{equation}
The results above are in good agreement on-shell with those in Ref.~\cite{Escribano:2010wt}, while they incorporate the missing $\Lambda_1$ terms in Ref.~\cite{Escribano:2010wt} that are necessary to obtain the scale-independent $\tilde{\Lambda}$ combination.
Concerning the non-pole contribution, we find\footnote{To LO precision, $\sin\phiFL\cos\phiFL(M_{\eta'}^2 -M_{\eta}^2) = (\sqrt{2}/3)M_0^2$, whereas $\cos^2\phiFL M_{\eta}^2 +\sin^2\phiFL M_{\eta'}^2 = M_{\pi}^2 +(2/3)M_0^2$, with $M_0^2$ the $\eta'$ mass in the chiral limit that, to LO accuracy, reads $M_0^2 =M_{\eta'}^2 +M_{\eta}^2 -2M_K^2$.} 
\begin{align}
    \mathcal{M}_{\textrm{non-pole}}^{\eta';s} &= 
    \frac{4\sqrt{2}B_0}{F_{\pi}^3}\sin^2\phi_{qs}\cos\phi_{qs}\big(M_{\eta'}^2 -M_{\eta}^2\big)(3L_2 +L_3),
    \notag \\
    \mathcal{M}_{\textrm{non-pole}}^{\eta';q} &= 
    \frac{2B_0}{F_{\pi}^3}\sin\phi_{qs}\Big[
    4(3L_2 +L_3)\big(s_{\ell} +\cos^2\phi_{qs}M_{\eta}^2 +\sin^2\phi_{qs}M_{\eta'}^2\big) +2M_{\pi}^2L_5
    \Big]
    \notag\\
    & -\frac{2B_0}{F_{\pi}} \sin\phi_{qs}  \left[ 
    1 - \frac{4L_5}{F_{\pi}^2}\big( M_{\eta'}^2 +4M_{\pi}^2 -s_{\ell}\big) +\frac{64 L_8 M_{\pi}^2}{F_{\pi}^2}
    \right] +\frac{2B_0}{F_{\pi}}\frac{\tilde{\Lambda}}{\sqrt{6}}\cos\theta_{81}, 
\end{align}
where the last line comes from genuine non-pole diagrams that are present, at NLO accuracy, only for the light-quark component.  Finally, the pseudoscalar matrix elements at NLO read
\begin{align}
    \bra{0} P^s \ket{P} &{}= c_P^s \sqrt{2}B_0F_{\pi}\left[ 1 -\frac{8L_5 M_K^2}{F_{\pi}^2} +\frac{16L_8(2M_K^2 -M_{\pi}^2)}{F_{\pi}^2} \right] -c_P^{\Lambda} \frac{ B_0 F_{\pi}}{\sqrt{6}}\tilde{\Lambda}, \notag\\
    \bra{0} P^q \ket{P} &{}= c_P^q 2B_0F_{\pi}\left[ 1 -\frac{8(L_5 -2L_8) M_{\pi}^2}{F_{\pi}^2} \right] -c_P^{\Lambda}\frac{ B_0 F_{\pi}}{2\sqrt{3}}\tilde{\Lambda},
\end{align}
where $c_{\eta}^q =c_{\eta'}^s = \cos\phi_{qs} $,  $c_{\eta'}^q =-c_{\eta}^s = \sin\phi_{qs} $, $c_{\eta'}^{\Lambda} = \cos\theta_{81}$, and $c_{\eta}^{\Lambda} = -\sin\theta_{81}$. We take the values for $L_5=1.66(23)\times10^{-3}$ and $L_8=1.08(13)\times10^{-3}$ from the lattice~\cite{largenclecs}. 
The remaining combination of LECs, $3L_2 +L_3$, is fixed based on 
the $\eta'\to\eta\pi^+\pi^-$ decay in \cref{sec:AppBSMetaPpolynomial}. Once more, a consistency check for the non-pole part is provided by the vanishing result of the strange-quark matrix element in the soft-pion limit discussed in \cref{sec:etaBSM}.

\subsubsection{Unitarization\label{sec:unit}}

Beyond the perturbative large-$N_c$ $\chi$PT calculation, the $\eta'$ decay requires to account for non-perturbative rescattering effects due to the relatively large available phase space.  It is, in fact, well known that the pure large-$N_c$ $\chi$PT representation of the closely related decay amplitudes $\eta'\to\eta\pi\pi$ fails to describe the available experimental data in a satisfactory manner~\cite{Escribano:2010wt}, such that some way of unitarization of the final-state interactions is phenomenologically required~\cite{Escribano:2010wt,Isken:2017dkw,Gonzalez-Solis:2018xnw,Akdag:2021efj}.\footnote{We consider analogous unitarization for the $S$-wave in the BSM amplitude for $\eta\to\pi^+\pi^-\mu^+\mu^-$ less pressing at this stage, given the smallness of the available phase space, as well as the sizable uncertainties induced by the NLO low-energy constants. Note that the one-loop representation of the strange matrix element already contains perturbative unitarization at leading order.}

The by far dominant rescattering effect at low energies occurs in the pion--pion isospin $I=0$ $S$-wave; the $D$-wave phase shift is known to be very small at the energies available.\footnote{Similarly, left-hand cuts due to $\pi\eta$ $S$-wave intermediate states are expected to be weak and suppressed at low energies~\cite{Kubis:2009sb,Albaladejo:2015aca,Lu:2020qeo}.}  
We therefore project our amplitude onto partial waves along the lines of Ref.~\cite{Escribano:2010wt}, which allows us to express the amplitude as 
\begin{equation}
    \mathcal{M}_{AB\to\pi\pi} = \mathcal{M}_{AB\to\pi\pi}^{l=0} +\mathcal{M}_{AB\to\pi\pi}^{l=2}, \qquad 
    \mathcal{M}_{\textrm{non-pole}}^{\eta'} = \mathcal{M}_{\textrm{non-pole}}^{\eta' l=0}
    , 
\end{equation}
where we reflect the fact that, due to the essentially polynomial nature of the large-$N_c$ amplitudes, no partial waves beyond $D$-waves occur at all, and the non-pole part is in fact a pure $S$-wave up to the order considered here.
The partial waves obtained for the scattering amplitude(s) in \cref{eq:etapiscatt} are found to be\footnote{Note that the last term in the first line of Eq.~(3.15) in Ref.~\cite{Escribano:2010wt}  should read $2m_{\eta'}^2m_{\pi}^2/s$ instead of $2m_{\eta'}^2m_{\eta}^2/s$.}
\begin{align}
    \mathcal{M}_{AB\to\pi\pi}^{l=0} &= c_{AB}\frac{M_{\pi}^2}{3F_{\pi}^2} +W_{AB}(s) \notag\\ &+ c_{AB}\frac{4(3L_2 +L_3)}{3F_{\pi}^4}\bigg[ \frac{2}{3}\lambda\beta_{\pi}^2  +\frac{2M_{\pi}^2(M_{\eta'}^2 -s_{\ell})^2}{s} +2(M_{\eta'}^2-M_{\pi}^2)(s_{\ell} -M_{\pi}^2) \bigg], \notag\\
    \mathcal{M}_{AB\to\pi\pi}^{l=2} &=  c_{AB}\frac{4(3L_2 +L_3)}{3F_{\pi}^4}\frac{\lambda\beta_{\pi}^2}{6}\big(3\cos^2\theta_{\pi} -1\big), \quad 
    \cos\theta_{\pi} = \frac{t-u}{\lambda^{1/2}(M_{\eta'}^2,s,s_{\ell})\beta_{\pi}(s)}.  \label{eq:Pwaves}
\end{align}
We incorporate $\pi\pi$ rescattering effects for the $S$-wave component only, upon substituting
\begin{equation}
    \mathcal{M}^{l=0} \to \mathcal{M}^{l=0} \, (1+\alpha s) \Omega_0^0(s), \qquad
    \Omega_0^0(s) = \exp \bigg( \frac{s}{\pi}\int_{4M_\pi^2}^{\infty}ds' \frac{\delta_0^0(s')}{s'(s'-s)} \bigg) , \label{eq:Omnes}
\end{equation}
with $\Omega_0^0(s)$ the corresponding Omn{\`e}s function~\cite{Omnes:1958hv} from Ref.~\cite{Isken:2017dkw}.  
The multiplicative linear polynomial $(1+\alpha s)$, with the free slope parameter $\alpha$, effectively accounts for the coupled-channel effects that set in above $K\bar K$ threshold, which are known to be strong in the $I=0$ $S$-wave and can impact the $\pi\pi$ final state very differently~\cite{Donoghue:1990xh,Moussallam:1999aq,DescotesGenon:2000ct,Daub:2012mu,Celis:2013xja}; as well as for potential left-hand cuts.
Since our approach contains the $\eta'\to\eta\pi\pi$ subamplitude, we can test the goodness of this unitarization approach by comparing our representation to Dalitz plot data for $\eta'\to\eta\pi^+\pi^-$~\cite{BESIII:2017djm}, which also allows us to fix the parameter $\alpha$ above.   This comparison is discussed in detail in \cref{sec:AppBSMetaPpolynomial}, where our unitarization method is observed to work very well.
We from now on assume $\alpha$ to be universal for all rescattering effects, \textit{i.e.}, that the value extracted from the $\eta'\to\eta\pi^+\pi^-$ Dalitz plot is applicable in \cref{eq:Omnes}.


\section{Results}
\label{sec:results}

Taking the description in \cref{sec:BSM} to feed the hadronic matrix element in \cref{eq:Asinphi}, we find the following result for the asymmetry\footnote{To illustrate the necessity of an NLO calculation for the strange quark, we remark that the LO result for the $\eta$ would read $A_{\phi}^{\eta} =2.5\times 10^{-5}\operatorname{Im}c_{\ell edq}^{2222}$ ---an order of magnitude larger than the central value of the full number in Eq.~\eqref{eq:Nasymmetries}; our final result arises from a cancellation of similar---but opposite---LO and NLO contributions. Regarding the $\eta'$, ignoring unitarization effects, the LO result would read $A_{\phi}^{\eta'} =3.5\times10^{-5}\big(\operatorname{Im}c^{(1)2211}_{\ell equ} +\operatorname{Im}c_{\ell edq}^{2211}\big) + 0.3\times10^{-5}\operatorname{Im}c_{\ell edq}^{2222}$, whereas including NLO corrections would result in $A_{\phi}^{\eta'} =6.2\times10^{-5}\big(\operatorname{Im}c^{(1)2211}_{\ell equ}+\operatorname{Im}c_{\ell edq}^{2211}\big) -2.7\times10^{-5}\operatorname{Im}c_{\ell edq}^{2222}$, which reflects the anticipated sizable NLO corrections in the $\eta'$ sector.} 
\begin{align}
    A_{\phi}^{\eta} &{}= 47(14)\times 10^{-5}\Big(\operatorname{Im}c^{(1)2211}_{\ell equ}+\operatorname{Im}c_{\ell edq}^{2211}\Big) -0.4(2.2)\times10^{-5}\operatorname{Im}c_{\ell edq}^{2222}, \notag\\
    A_{\phi}^{\eta'} &{} =2.9(5)\times10^{-5}\Big(\operatorname{Im}c^{(1)2211}_{\ell equ}+\operatorname{Im}c_{\ell edq}^{2211}\Big) -1.4(5)\times10^{-5}\operatorname{Im}c_{\ell edq}^{2222}, \label{eq:Nasymmetries}
\end{align}
where the SM part in the ratio has been taken from \cref{tab:SM-BRs}.
We note that the Wilson coefficients above should be taken at a scale $\mu = 2\GeV$ in order to match that of $B_0$ and render the result scale-independent.
The error for the $\eta$ case is dominated, for the strange-quark contribution, by $L_6^r$ and $3L_{2}^r +L_3^r$, contributing to the total uncertainty as $1.5\times10^{-5}\operatorname{Im}c_{\ell edq}^{2222}$ and $1.2\times10^{-5}\operatorname{Im}c_{\ell edq}^{2222}$, respectively.
For the $\eta$ light-quark matrix element, which we have only evaluated at LO, we assume a natural convergence behavior of the $SU(3)$ expansion and add a 30\% uncertainty by hand.
For the $\eta'$, the errors are harder to assess. Theory uncertainties not yet covered consist of next-to-next-to-leading-order (NNLO) corrections; note however that our unitarization scheme already includes potentially large rescattering contributions in full, and we would expect parts of the remaining NNLO contributions to be absorbed in a re-fit of the $\eta'\to\eta\pi\pi$ Dalitz plot, \textit{cf.}\ \cref{sec:AppBSMetaPpolynomial}.  
The main effect at higher orders will therefore consist in corrections to the strict relation between the $\eta$-pole contribution (rigorously fixed by $\eta'\to\eta\pi\pi$) and the remaining parts of the amplitude.  We guesstimate this by taking the size of the NLO amplitude and divide it by $N_c$.  This leads to the uncertainties quoted in Eq.~\eqref{eq:Nasymmetries}.
The above results can be used to find the sensitivity that can be reached at a given experiment, which is limited by the precision that can be achieved for the $A_{\phi}$ asymmetry. In the following, we estimate the sensitivities at the proposed REDTOP facility, which would become the largest $\eta/\eta'$ factory in the future. To do so, we make a poor-theorist estimate assuming that all backgrounds arise from statistical fluctuations of the SM process itself, thus neglecting other backgrounds that would require a dedicated experimental analysis. This yields $\Delta A_{\phi} = 1/\sqrt{N}$, where $N$ is the total number of events. Taking $N_{\eta}=5\times 10^{12}$, $N_{\eta'}=4\times 10^{10}$,\footnote{These numbers include reconstruction efficiencies of $5\%$ based on $\eta^{(\prime)}\to2\mu^+2\mu^-$ studies and input from Table~XXIII of Ref.~\cite{REDTOP:2022slw}. We acknowledge C.~Gatto for discussions on this point.} and considering a single Wilson coefficient at a time, we find the results in \cref{table:Sensitivityresults2}, which for completeness also shows the results from previous studies for different channels~\cite{Sanchez-Puertas:2018tnp,Sanchez-Puertas:2019qwm,Escribano:2022zgm}.
\begin{table}
	\centering
	\def\arraystretch{1.3}
	\begin{tabular}{ccccc}
		\toprule
		 Process & Asymmetry & $\operatorname{Im}c_{\ell edq}^{2222}$ & $\operatorname{Im}c^{(1)2211}_{\ell equ}$ & $\operatorname{Im}c_{\ell edq}^{2211}$ \\ \midrule
         $\eta\to\pi^+\pi^-\mu^+\mu^-$ & $A_{\phi}$ & $1584$ & $12$ & $12$\\ 
         $\eta^{\prime}\to\pi^+\pi^-\mu^+\mu^-$ & $A_{\phi}$ & $77$ & $36$ & $36$\\ \midrule
		 $\eta\to\pi^0\mu^+\mu^-$ & $A_L$ & $0.7$ & $0.07$ & $0.07$\\
		 $\eta'\to\pi^0\mu^+\mu^-$ & $A_L$ & $11$ & $2.4$ & $2.5$\\
		 $\eta'\to\eta\mu^+\mu^-$ & $A_L$ & $5$ & $68$ & $79$\\ \midrule
		 $\eta\to\mu^+\mu^-$ & $A_T$ & $0.005$ & $0.007$ & $0.007$\\ \midrule
         nEDM & --- & $\leq 0.02$ & $\leq 0.001$ & $\leq 0.002$\\
         \bottomrule
	\end{tabular}
\caption{Results for the REDTOP sensitivities to the Wilson coefficients associated with $CP$-violating SMEFT operators. For completeness we also show the sensitivities from other leptonic and semileptonic channels from Refs.~\cite{Sanchez-Puertas:2018tnp,Sanchez-Puertas:2019qwm,Escribano:2022zgm}.}
\label{table:Sensitivityresults2}
\end{table}
As a result, we find that the precision that can be achieved for $\etap\to\pi^+\pi^-\mu^+\mu^-$ at REDTOP is not competitive with the bounds set by the nEDM, which are only beaten by the $\eta\to\mu^+\mu^-$ channel and for the strange-quark operator. 

At this stage one might wonder about the reasons behind the inferior sensitivity for $\operatorname{Im}c_{\ell edq}^{2222}$ for the $\eta$ channel in particular as compared to $\eta\to\pi^0\mu^+\mu^-$, despite both processes having similar branching ratios within the SM and with the latter having a BSM contribution suppressed by isospin breaking (IB). This, together with the fact that muon polarimetry is not required in $A_\phi$, was the main point that made the present study such an appealing case. To understand this, we note that $\eta\to\pi^+\pi^-\mu^+\mu^-$ is phase-space suppressed, which equally affects the BSM part, while the $\eta\to\pi^0\mu^+\mu^-$ suppression comes mainly from a factor of $(\alpha/\pi)^2$---in the SM, the decay proceeds dominantly via two-photon intermediate states~\cite{Escribano:2020rfs}. Second, there is a lepton chiral suppression $\mathcal{O}(m_{\mu}/M_{\eta})$ in this process (\textit{cf.}\ the discussion in \cref{sec:main}). Third, we have seen that the matrix element is chirally suppressed at LO as $M_{\pi}^2/M_{\eta}^2$. Finally, the LO and NLO contributions lead to an accidental order-of-magnitude cancellation, so that we might estimate a suppression of the sensitivity in $\eta\to\pi^+\pi^-\mu^+\mu^-$ relative to the one in $\eta\to\pi^0\mu^+\mu^-$ according to \begin{equation}
    \bigg(\frac{M_{\pi}}{M_{\eta}}\bigg)_{\chi_{SU(3)}}^2 \bigg(\frac{m_{\mu}}{M_{\eta}}\bigg)_{\chi_{\ell}}\Big(\frac{1}{10}\Big)_{\textrm{NLO}} 100_{\textrm{IB}}\Big(\frac{\alpha}{\pi}\Big) \sim \frac{1}{3500} ,
\end{equation}
which roughly explains the difference. 
Regarding the $\eta'$ we find the current sensitivities an order-of-magnitude lower than in $\eta'\to\pi^0\mu^+\mu^-$ (but comparable to $\eta'\to\eta\mu^+\mu^-$), once more due to the $(\alpha/\pi)$-enhancement of the latter, which in the end seems to compensate the lower SM branching ratio and isospin suppression---a feature that would have been hard to anticipate without the current study.

This work complements the previous effort in Refs.~\cite{Sanchez-Puertas:2018tnp,Sanchez-Puertas:2019qwm,Escribano:2022zgm} in the context of $CP$-violation searches in leptonic and semileptonic $\eta$ and $\eta'$ decays within the SMEFT framework. Within such framework, and considering REDTOP as the largest $\eta$ factory in the future, the $\eta\to\mu^+\mu^-$ decay stands out as the single channel competitive with nEDM constraints, though muon polarimetry is required. 


\section{Summary}
\label{sec:summary}

With the advent of future $\etap$ factories aiming to find physics beyond the Standard Model via discrete-symmetry tests, it is timely to assess the physics within reach to help guiding the experimental programs. In this article, we extended the study of $P$- and $CP$-violating $\etap$ semileptonic decays initiated in Refs.~\cite{Sanchez-Puertas:2018tnp,Sanchez-Puertas:2019qwm,Escribano:2022zgm} to $\etap\to\pi^+\pi^-\mu^+\mu^-$. $CP$ violation in these processes is severely constrained by the nEDM, which we have used to place bounds using SMEFT. Within this framework, the most relevant contributions arise from scalar quark--lepton dimension-6 Fermi operators, whose nEDM contribution appears only at two loops. 
Interestingly enough, these operators induce a $CP$-violating $\sin\phi_{\pi\ell}$ asymmetry, which should be incorporated in experimental analyses that commonly target a $\sin2\phi_{\pi\ell}$ asymmetry instead. The latter is motivated in scenarios with $CP$-violating form factors that, we argued, suffer from yet stronger bounds from the nEDM or require higher-dimensional operators in SMEFT. 

In our study, we worked out the corresponding hadronic matrix elements appearing in $\etap\to\pi^+\pi^-\mu^+\mu^-$ using $\chi$PT and standard unitarization techniques; such matrix elements might be also interesting for other BSM scenarios. As a result, we find moderate sensitivities that probably cannot overcome nEDM bounds. This is mostly due to the small phase space available for the $\eta$, as well as chiral suppression and accidental cancellations for the strange-quark contribution. The results are compared to other channels, which should help guiding the experimentalists targeting the most interesting decays. In particular, the $\eta\to\mu^+\mu^-$ decay continues to be the most promising case at future $\etap$ factories, provided muon polarization can be assessed. This mostly exhausts the list of semileptonic $\etap$ decays, with the possible exception of $\etap\to\pi^0\pi^0\mu^+\mu^-$, which might be interesting to consider in the future.

\acknowledgments
\begin{sloppypar}
We are grateful to Hakan Akdag and Tobias Isken for providing us with parameterizations for phase shifts and Omnès functions, as well as for useful discussions.  We thank Martin Hoferichter for helpful comments on the manuscript.
Financial support was provided by the Swiss National Science Foundation
under Project No.\ PCEFP2\_181117, by the DFG (CRC 110, ``Symmetries and the Emergence of Structure in QCD''), the European Union's Horizon 2020 Research and Innovation Programme (grants no.\ 754510 [EU, H2020-MSCA-COFUND2016] and no.\ 824093 [H2020-INFRAIA-2018-1]), the Spanish MINECO, MCIN/AEI/10.13039/501100011033 (grants PID2020-112965GB-I00 and PID2020-114767GB-I00), and Junta de Andaluc{\'i}a (grants POSTDOC\_21\_00136 and FQM-225).
\end{sloppypar}


\appendix


\section{Phase space description \label{sec:AppPhaseSpace}}

In the following, we adopt the Cabibbo--Maksymowicz~\cite{Cabibbo:1965zzb}
description for the four-body phase space, see Ref.~\cite{Kampf:2018wau} for details. In particular, the phase space can be described as sequential two-body decays. Labeling the momenta and invariants as in \cref{sec:main},
\begin{align}
d\Pi_{4} =&\dfrac{\lambda^{1/2}}{2^{14}\pi^{6}M_{\eta^{(\prime)}}^2} ds\, ds_{\ell}\, \beta_{\pi}\,d\cos\theta_{\pi}\, \beta_{\ell}\,d\cos\theta_{\ell}\, d\phiKin\, ,
\end{align}
where 
in addition we introduced $\theta_{\ell}$ as the angle in the dilepton frame (see details in Ref.~\cite{Kampf:2018wau}). 
The integration limits are given as $s \in [4M_{\pi}^2 , (M_{\eta^{(\prime)}} -2m_{\ell})^2]$, $s_{\ell} \in [ 4m_{\ell}^2, (M_{\eta^{(\prime)}} -\sqrt{s})^2 ]$, $\cos\theta_{\pi,\ell} \in[-1,1]$, and $\phiKin\in[0,2\pi]$.
This way one can express all the relevant invariants as follows (we use $p_{ij} =p_i +p_j$, $\bar{p}_{ij} =p_i -p_j$):
\begin{align}
    2p_{12}\cdot p_{34} &{}= M_{\eta^{(\prime)}}^2 -s -s_{\ell} \, , \quad 
    2\bar{p}_{12}\cdot p_{34} = \lambda^{1/2}\beta_{\pi}\cos\theta_{\pi},\, \quad 
    2\bar{p}_{34}\cdot p_{12} = \lambda^{1/2}\beta_{\ell}\cos\theta_{\ell}\, , \nonumber\\
    2\bar{p_{12}}\cdot\bar{p}_{34} &{}= (M_{\eta^{(\prime)}}^2 -s -s_{\ell})\beta_{\pi}\beta_{\ell}\cos\theta_{\pi}\cos\theta_{\ell} -2\sqrt{s\, s_{\ell}}\beta_{\pi}\beta_{\ell}\sin\theta_{\pi}\sin\theta_{\ell}\cos\phiKin\, , \nonumber \\
    \epsilon_{\mu\nu\rho\sigma}p_{1}^{\mu}p_{2}^{\nu}p_{3}^{\rho}p_{4}^{\sigma} &{}=  -\frac{\lambda^{1/2} \sqrt{s\, s_{\ell}}}{8} \beta_{\pi}\beta_{\ell}\sin\theta_{\pi}\sin\theta_{\ell}\sin\phiKin \, ,
\end{align}


\section{BSM amplitude for the \texorpdfstring{$\eta$}{eta} meson at NLO \label{sec:AppBSMeta}}

The single-variable functions $W_{\eta\eta}(s)$ and $U_{\eta\eta}(t)$ appearing in the decomposition of the scattering amplitude read 
\begin{align}
	W_{\eta\eta}(s) =
	\dfrac{1}{F_{\pi}^4}&\Bigg\{\left(s-2M_{\pi}^2\right)\left(s-2M_{\eta}^2\right)\bigg[4\bigg(2L_1^r+\dfrac{L_3^r}{3}\bigg)-\dfrac{3}{8}\dfrac{1}{16\pi^2}\left(1+L_K\right)\bigg] \notag\\&
	+M_{\pi}^2M_{\eta}^2\bigg[32\bigg(-L_7^r+L_6^r-\dfrac{1}{6}L_5^r-L_4^r\bigg)+\dfrac{1}{16\pi^2}\left(\dfrac{23}{18}+2L_K-\dfrac{2}{9}L_{\eta}\right)\bigg]  \notag\\&
	+M_{\pi}^4\bigg[16L_8^r+32L_7^r+\dfrac{1}{16\pi^2}\left(-\dfrac{1}{9}-\dfrac{2}{9}R_{\pi\eta}-\dfrac{1}{6}L_K-\dfrac{1}{6}L_{\eta}-\dfrac{1}{2}L_{\pi}\right)\bigg]  \notag\\&
	+s\Sigma_{\eta\pi}\bigg[8L_4^r-\dfrac{1}{2}\dfrac{1}{16\pi^2}(1+L_K)\bigg]+sM_{\pi}^2\bigg[\dfrac{1}{3}\dfrac{1}{16\pi^2}\log(\dfrac{M_K^2}{M_{\pi}^2})\bigg]\notag\\&
	-\dfrac{1}{6}\bar{J}_{\pi\pi}(s)M_{\pi}^2(M_{\pi}^2-2s) +\dfrac{1}{54}\bar{J}_{\eta\eta}(s)M_{\pi}^2(16M_K^2-7M_{\pi}^2) 
	\notag\\ &
	-\dfrac{1}{24}\bar{J}_{KK}(s)s(8M_K^2-9s) 
	+2\Sigma_{\eta\pi}\left(s- \Sigma_{\eta\pi} \right)\bigg[4\left(L_2^r+\dfrac{L_3^r}{3}\right)-\dfrac{3}{8}\dfrac{1+L_K}{16\pi^2}\bigg]  \notag\\&
	+\dfrac{2\Sigma_{\eta\pi} -s}{288\pi^2}\bigg[2M_K^2+M_{\pi}^4\dfrac{\Sigma_{\eta\pi} -2M_{\eta}^2R_{\pi\eta}}{(M_{\pi}^2-M_{\eta}^2)^2}\bigg] 
	\Bigg\}\, ,
\end{align}
which agrees with the expression in Ref.~\cite{Albaladejo:2015aca} except for the last line (that originates from our condition $U(t)=U'(t)=0$), and
\begin{align}
	U_{\eta\eta}(t) = 
	\dfrac{1}{F_{\pi}^4}&\Bigg\{t^2\bigg[4\Big(L_2^r+\dfrac{L_3^r}{3}\Big)-\dfrac{3}{8}\dfrac{1}{16\pi^2}(1+L_K)\bigg]  
	-\dfrac{t}{288\pi^2}\bigg[2M_K^2+M_{\pi}^4\dfrac{\Sigma_{\eta\pi} -2M_{\eta}^2R_{\pi\eta}}{(M_{\pi}^2-M_{\eta}^2)^2}\bigg] \notag\\
	& +\dfrac{1}{9}M_{\pi}^4\bar{J}_{\pi\eta}(t)+\dfrac{1}{24}\left(4M_{K}^2-3t\right)^2\bar{J}_{KK}(t)\Bigg\} \,,
\end{align}
which also agrees with Ref.~\cite{Albaladejo:2015aca} once removing the $U(0)$ and $U'(0)$ terms.
The matrix element $\matrixel{0}{\bar{s} i \gamma_{5} s}{\eta}$ appearing in the pole part is given by
\begin{align}
	\matrixel{0}{\bar{s} i \gamma^{5} s}{\eta} =   
	-\dfrac{2}{\sqrt{3}}F_{\pi}B_0 & \bigg[ 1 
	+ \frac{M_{\pi}^2 L_{\pi}}{16\pi^2F_{\pi}^2} 
	- \frac{M_{\eta}^2 L_{\eta}}{24\pi^2F_{\pi}^2} 
	-\dfrac{4 M_{\eta}^2}{F_{\pi}^2}\big(3L_{4}^{r}+L_{5}^{r}-6L_{6}^{r}-6L_{7}^{r}-6L_{8}^{r}\big) \notag\\ & 
	-\dfrac{4 M_{\pi}^2}{F_{\pi}^2}\big(3L_{4}^{r}+L_{5}^{r}-6L_{6}^{r}+6L_{7}^{r}+2L_{8}^{r}\big) \bigg]. 
\end{align}
Finally, the single-variable functions appearing in the non-pole part are given by 
\begin{align}
W^{\eta;s}_{\text{non-pole}}(s,s_\ell)= \frac{2 B_0F_{\pi}}{\sqrt{3}F_{\pi}^4} &\Bigg\{\frac{33 D_{\eta}+72 M_{\eta}^2+52 M_{\pi}^2-42s}{1152 \pi ^2} 
+ M_{\pi}^4\dfrac{\Sigma_{\eta\pi} -2M_{\eta}^2R_{\pi\eta}}{288\pi^2(M_{\pi}^2-M_{\eta}^2)^2} \notag\\
& +\frac{s \bar{J}_{KK}(s)}{8}  
+\frac{2 M_{\pi}^2 \bar{J}_{\eta\eta}(s)}{9} 
+4 s \left(L_2^r -2 L_1^r +2 L_4^r -\dfrac{L_K}{128\pi^2}\right) \notag\\ & +4M_{\pi}^2
\left(4 L_1^r+\frac{2
	L_3^r}{3}-6 L_4^r-\frac{L_5^r}{3}+8 L_6^r-4
L_7^r-\dfrac{L_{\eta}}{288\pi^2}\right)\notag\\
&-4\left(D_{\eta}+ 2\Sigma_{\eta\pi}\right) \left(L_2^r+\frac{L_3^r}{3}-\dfrac{3L_K}{512\pi^2}\right)  
\Bigg\}
\end{align}
and 
\begin{equation}
U^{\eta;s}_{\text{non-pole}}(t)=\dfrac{B_0F_{\pi}}{4\sqrt{3}F_{\pi}^4}\bigg\{\left(4 M_{K}^2-3 t\right) \bar{J}_{KK}(t)-\frac{t}{24\pi^2}\bigg\}\, ,
\end{equation}
where, following Refs.~\cite{Albaladejo:2015aca,GomezNicola:2001as}, we have introduced
\begin{equation}
L_{P} = \log(\dfrac{M_{P}^2}{\mu^2}), \qquad R_{PQ}=\dfrac{M_{P}^2\log(M_{P}^2/M_{Q}^2)}{M_{P}^2-M_{Q}^2}, \qquad D_{\eta}=s_{\ell}-M_{\eta}^2\, ,
\end{equation}
with $\mu$ the renormalization scale, which is conventionally taken as $\mu =0.77\GeV$~\cite{Scherer:2012xha,Bijnens:2014lea}. We note that the combinations in brackets and parenthesis in the equations above are indeed scale-invariant; the $\beta$-functions determining the scale dependence of the $L_i^r$ can be found in Ref.~\cite{Gasser:1984gg}. The values used for the LECs at such reference scale are given in \cref{table:SU3LECS}.
\begin{table}
    \centering
    \begin{tabular}{cccccccc} \toprule
    $L_1^r$ & $L_2^r$ & $L_3^r$ & $L_4^r$ & $L_5^r$ & $L_6^r$ & $L_7^r$ & $L_8^r$  \\ \midrule
    $1.0(1)$ & $1.6(2)$ & $-3.8(3)$ & $0.0(3)$ & $1.2(1)$ & $0.0(4)$ & $-0.3(2)$ & $0.5(2)$  \\\bottomrule
    \end{tabular}
    \caption{Values for the LECs (in units of $10^{-3}$) taken from Ref.~\cite{Bijnens:2014lea}. We choose the results of the combined fit at NLO for the range $-0.3\times 10^{3}\leq L_4^{r}\leq 0.3\times 10^{3}$. The renormalization scale is $\mu=0.77\,\text{GeV}$. 
    }
    \label{table:SU3LECS}
\end{table}

Finally, $\bar{J}_{PQ}(s)= J_{PQ}(s) -J_{PQ}(0)$ is related to the standard scalar two-point function as $16\pi^2 J_{PQ}(s) = B_0(s,M_P^2,M_Q^2)$. Its particular expression reads~\cite{Gasser:1984gg}
\begin{align}\label{eq:B0integral}
	\bar{J}_{PQ}(s) =&\, \dfrac{1}{32\pi^2}\Bigg[2+\left(\dfrac{\Delta_{PQ}}{s}-\dfrac{\Sigma_{PQ}}{\Delta_{PQ}}\right)\log\bigg(\dfrac{M_{Q}^2}{M_{P}^2}\bigg) 
	-\dfrac{\nu(s)}{s}\log(\dfrac{(s+\nu(s))^2-\Delta_{PQ}^2}{(s-\nu(s))^2-\Delta_{PQ}^2})\Bigg]\, ,
\end{align}
where $\Sigma_{PQ} = M_{P}^2+M_{Q}^2$, $\Delta_{PQ} = M_{P}^2-M_{Q}^2$, and $\nu^2(s) = \lambda(s,M_P^2, M_Q^2)$. For equal masses, the expression simplifies considerably to 
\begin{equation}
    \bar{J}_{PP}(s) = \dfrac{1}{16\pi^2}\bigg[2+\beta_P(s)\log(\dfrac{\beta_P(s)-1}{\beta_P(s)+1})\bigg].
\end{equation}


\section{Phenomenology of \texorpdfstring{$\eta'\to\eta\pi\pi$}{eta'} \label{sec:AppBSMetaPpolynomial}}

The scattering amplitude $\mathcal{M}_{\eta'\eta\to\pi\pi}$ that is part of the $\eta$-pole contribution to the matrix element in Eq.~\eqref{eq:EtaPAmp} can be related to the decay amplitude $\mathcal{M}_{\eta'\to\eta\pi\pi}$ via crossing symmetry. However, it was shown that large-$N_c$ $\chi$PT to NLO is not able to describe the decay $\eta'\to\eta\pi\pi$ properly~\cite{Escribano:2010wt,Isken:2017dkw}. 
If the pseudoscalar matrix element in \cref{eq:EtaPAmp} were dominated by the $\eta$ pole, as one may have naively expected, one possibility might have been to employ the dispersion-theoretical amplitude representations for $\eta'\to\eta\pi\pi$~\cite{Isken:2017dkw,Akdag:2021efj} directly.
However, since the asymmetry involves the interference of the BSM and the SM contributions, with the latter lacking an $\eta$ pole, the asymmetry contains a single propagator $(s_\ell -M_{\eta}^2 +iM_{\eta}\Gamma_{\eta})^{-1}$ rather than its square. Consequently, both terms in its decomposition into principal value and imaginary part survive, and both pole and non-pole contributions must be taken into account.  Furthermore, in order for the amplitude to fulfill the Adler consistency conditions, all its parts---$\eta$ and $\eta'$ pole terms as well as the non-pole contribution---need to be calculated consistently within the same framework.  For this reason, we need to develop a somewhat simplified unitarization scheme that can be applied throughout.

The simplest possibility to account for non-perturbative $\pi\pi$ $S$-wave rescattering is to multiply the $S$-wave component with the corresponding Omn\`es function $\Omega_{0}^{0}(s)$,
\begin{equation}
    \mathcal{M}^{l=0} \to \mathcal{M}^{l=0} \, \Omega_0^0(s).
\end{equation}
However, as we will see below, this representation does not describe the $\eta'\to\eta\pi\pi$ Dalitz plot data well enough. 
In order to improve the phenomenology and provide a little more flexibility, we furthermore multiply the Omn\`es factor by
a linear polynomial $P(s)=1+\alpha s$ according to
\begin{equation}
    \mathcal{M}^{l=0} \to \mathcal{M}^{l=0} \, \Omega_0^0(s)P(s)=\mathcal{M}^{l=0} \, \Omega_0^0(s)(1+\alpha s).
\end{equation}
The linear polynomial has to fulfill $P(s=0)=1$ in order to preserve the soft-pion theorem, see \cref{sec:etaBSM}. The additional free parameter $\alpha$ is determined from a fit to the Dalitz plot distribution. 
The physical motivation to multiply the $S$-wave component not only with the Omn\`es function but also with a linear polynomial is to take potentially large inelastic effects of $K\bar{K}$ intermediate states in the $I=0$ $S$-wave into account, and probably also allow for left-hand cut contributions to some extent. 

It is desirable to obtain an estimation about the goodness of the different models.  
The LECs in $\mathcal{M}^{l=0}$ as well as the parameter $\alpha$ are taken as constants that have to be fitted to data. 
As the low-energy constants $L_5$, $L_8$, and $\tilde{\Lambda}$ come with a factor $M_{\pi}^2$ and are therefore suppressed, we only adjust the combination $(3L_2+L_3)$ and the slope $\alpha$ to the decay rate $\Gamma_{\eta^{\prime}\rightarrow\eta\pi^{+}\pi^{-}}^{\textrm{PDG}}$ and the Dalitz plot distribution according to the recent BESIII data~\cite{BESIII:2017djm}. The remaining low-energy constants have been fixed from the literature~\cite{largenclecs}, see \cref{sec:etapBSM}. For the combined fit to the decay rate $\Gamma_{\eta^{\prime}\rightarrow\eta\pi^{+}\pi^{-}}^{\textrm{PDG}}$ and the BESIII data, we 
minimize
\begin{equation}
    \chi^2 = \sum_{i} \left(\dfrac{N\times|\mathcal{M}_{\eta'\to\eta\pi\pi}(X_i,Y_i)|^2-N_{\textrm{BESIII}}(X_i,Y_i)}{\Delta N_{\textrm{BESIII}}(X_i,Y_i)}\right)^2+\left(\dfrac{\Gamma_{\eta^{\prime}\rightarrow\eta\pi^{+}\pi^{-}}-\Gamma_{\eta^{\prime}\rightarrow\eta\pi^{+}\pi^{-}}^{\textrm{PDG}}}{\Delta \Gamma_{\eta^{\prime}\rightarrow\eta\pi^{+}\pi^{-}}^{\textrm{PDG}}}\right)^2,
\end{equation}
with $N$ being a normalization factor taking care of the fact that BESIII data is given in arbitrary units. 
We use $\Gamma_{\eta^{\prime}\rightarrow\eta\pi^{+}\pi^{-}}^{\textrm{PDG}}=79.9(2.7)\keV$~\cite{pdgphysics}.
Moreover, the decay rate $\Gamma_{\eta^{\prime}\rightarrow\eta\pi^{+}\pi^{-}}$ in terms of the squared matrix element $|\mathcal{M}_{\eta'\to\eta\pi\pi}|^2$ is given by
\begin{align}
\Gamma_{\eta'\rightarrow \eta \pi^{+}\pi^{-}}=&
\dfrac{1}{512\pi^3M_{\eta'}^3}\int ds\int d\cos(\theta_{\pi})\,\beta_\pi\lambda^{1/2}(s,M_{\eta'}^2,M_{\eta}^2) |M_{\eta'\rightarrow\eta\pi^{+}\pi^{-}}|^2 .
\end{align}
The minimization of $\chi^2$ has been performed upon restricting $3L_2+L_3$ to positive values, as it should not deviate too much from the literature values; in the limit $M_{\pi}\rightarrow 0$, the decay amplitude $M_{\eta'\rightarrow\eta\pi^{+}\pi^{-}}$ is strictly proportional to $3L_2+L_3$ and therefore both decay rate and Dalitz plot are insensitive to the sign.

After fixing the fit parameters through minimization of $\chi^2$, the Dalitz plot parameters can be obtained via the expansion of the squared amplitude around the center according to~\cite{Dalitz:1953cp,Fabri:1954zz,Weinberg:1960zza}
\begin{equation}\label{eq:Dalitzslope}
|\mathcal{M}_{\eta'\to\eta\pi\pi}|^2=|\mathcal{N}|^2[1+aY+bY^2+cX+dX^2]  ,
\end{equation}
where $|\mathcal{N}|^2$ is a normalization factor and the variables $X$ and $Y$ are defined according to
\begin{equation}
    X = \dfrac{\sqrt{3}}{Q}\left(T_{\pi^{+}}-T_{\pi^{-}}\right),\qquad Y=\dfrac{M_{\eta}+2M_{\pi}}{M_{\pi}}\dfrac{T_{\eta}}{Q}-1,
\end{equation}
with 
\begin{equation}
T_{\eta}=\dfrac{(M_{\eta'}-M_{\eta})^2-s}{2M_{\eta^{\prime}}}, \quad T_{\pi^{+}}=\dfrac{(M_{\eta'}-M_{\pi})^2-t}{2M_{\eta^{\prime}}}, \quad T_{\pi^{-}}=\dfrac{(M_{\eta'}-M_{\pi})^2-u}{2M_{\eta^{\prime}}} ,
\end{equation}
and $Q=T_{\eta}+T_{\pi^{+}}+T_{\pi^{-}}=M_{\eta'}-M_{\eta}-2M_{\pi}$. In the following, we set $c=0$ as odd terms in the Dalitz plot variable $X$ are forbidden due to charge conjugation symmetry~\cite{Akdag:2021efj}. The resulting values for the three different models can be found in \cref{table:Fitparameters}. 
\begin{table}
	\centering
	\def\arraystretch{1.3}
	\begin{tabular}{ccccc}
		\toprule
		 &$\mathcal{M}^{l=0}$ &$\mathcal{M}^{l=0} \, \Omega_0^0(s)$&$\mathcal{M}^{l=0} \, \Omega_0^0(s)P(s)$ & Ref.~\cite{BESIII:2017djm} \\ \midrule
		$N$ & $2.38(7)\times10^{-3}$ & $1.96(5)\times10^{-3}$ & $2.88(10)\times10^{-3}$ \\
		$3L_2+L_3$ & $1.06(1)\times10^{-3}$ & $0.74(1)\times10^{-3}$ & $0.98(2)\times10^{-3}$ \\ 
		$\alpha$ & $-$ & $-$ & $-3.05(2)\,\textrm{GeV}^{-2}$ \\ 
		$\chi^2/\textrm{d.o.f.}$ & $1.409$ & $2.282$ & $1.005$ \\ 
		$a$ & $-0.288$ & $-0.480$ & $-0.050(4)$ & $-0.056(4)(2)$ \\ 
		$b$ & $-0.000$ 
		& $\phantom{+}0.046$ & $-0.124(1)$ & $-0.049(6)(6)$ \\ 
		$d$ & $-0.081$ & $-0.051$ & $-0.0793(3)$ & $-0.063(4)(3)$\\ \bottomrule
	\end{tabular}
\caption{Results for the combined fit for all three models. Presented are the fit parameters as outcome of the minimization routine of $\chi^2$ and the respective Dalitz slope parameters.}
\label{table:Fitparameters}
\end{table}
The number or data points required to calculate the reduced $\chi^2$ is given by $10795$ bins in the Dalitz plot of Ref.~\cite{BESIII:2017djm}, plus the partial width.

We see from \cref{table:Fitparameters} that multiplying the $S$-wave component with the relevant Omn\`es function times a linear polynomial is the best model to describe the decay $\eta'\to\eta\pi\pi$. The respective $\chi^2$ suggests a very good fit that describes the data very well. Moreover, as it was pointed out in Ref.~\cite{Kaiser:2000gs}, we can compare the resulting value for $3L_2+L_3$ to the corresponding $SU(3)$ value, suggesting good agreement, see \cref{table:SU3LECS}. However, the value for $\alpha$ is large (and negative) and modifies the distribution even more strongly than the Omn\`es function $\Omega^{0}_{0}(s)$. Hence the modification through the linear polynomial should be seen as the introduction of an additional phenomenological fit parameter. We note that the polynomial introduces a zero in the $S$-wave for $s\sim 0.33\,\textrm{GeV}^{2}$; as the available phase space only goes up to $s\sim 0.17\,\textrm{GeV}^{2}$, this zero is located outside of the physical decay region.  A similar observation can be made in more sophisticated amplitude analyses of this decay~\cite{IskenStoffer:2022}.

The resulting values for the slope parameters also point towards the need to include the linear polynomial for the $S$-wave component. They can be compared to the values from Refs.~\cite{BESIII:2017djm,Isken:2017dkw}, only showing a slight remaining tension with the value for the parameter $b$. Note that the Dalitz plot parameters corresponding to the models without the parameter $\alpha$ are fixed theoretically when neglecting $M_{\pi}^2$-suppressed terms, as the fitting procedure only adjusts the combination $3L_2+L_3$ related to the normalization of the branching ratio.

\bibliographystyle{JHEP_mod}
\bibliography{base}

\end{document}